\documentclass[sigconf,nonacm]{acmart}

\usepackage{booktabs}
\title{Schadenfreude in the Digital Public Sphere: A Cross-National, Decade-Long Analysis of Facebook News Engagement}

\author{Nouar AlDahoul}
\authornote{Both authors contributed equally to this work.}
\affiliation{\institution{New York University Abu Dhabi}\city{Abu Dhabi}\country{UAE}}
\author{Hazem Ibrahim}
\authornotemark[1]
\affiliation{\institution{New York University Abu Dhabi}\city{Abu Dhabi}\country{UAE}}
\author{Majd Mahmutoglu}
\affiliation{\institution{Y{\i}ld{\i}z Technical University}\city{Istanbul}\country{T\"urkiye}}
\author{Hajra Tarar}
\affiliation{\institution{Lahore University of Management Sciences}\city{Lahore}\country{Pakistan}}
\author{Muhammad Fareed Zaffar}
\affiliation{\institution{Lahore University of Management Sciences}\city{Lahore}\country{Pakistan}}
\author{Talal Rahwan}
\authornote{Corresponding authors.}
\email{talal.rahwan@nyu.edu}
\affiliation{\institution{New York University Abu Dhabi}\city{Abu Dhabi}\country{UAE}}
\author{Yasir Zaki}
\authornotemark[2]
\email{yasir.zaki@nyu.edu}
\affiliation{\institution{New York University Abu Dhabi}\city{Abu Dhabi}\country{UAE}}

\begin{document}

\begin{abstract}
Pleasure at the misfortune of others, or schadenfreude, is widely believed to have become a routine feature of online news audiences, yet its prevalence and social patterning have rarely been measured at scale. Here, we examine ten years (2015--2024) of Facebook engagement with misfortune news across nine major outlets spanning the political spectrum in the United States, the United Kingdom, and India. We separate schadenfreude, in which a commenter expresses pleasure at the misfortune, from hostility without pleasure, using an ensemble of three open-weight language models validated against human annotation, and classify 1.06 million post--comment pairs. We find that schadenfreude constitutes 26\% of comments on news covering misfortune, a third more than the rate implied by ``Haha'' reactions on the same posts, and that the two measures agree at the post level (Spearman $\rho = 0.43$) while hostility does not ($\rho = 0.07$). Schadenfreude varies by less than one percentage point across the three countries and concentrates where theories of deserved suffering predict: on misfortunes framed as self-inflicted, on scandals and defeats rather than deaths, and on public figures rather than ordinary people. Hostility, in contrast, varies widely across outlets and responds to politics. Audiences of a left-leaning US outlet became 5.5 percentage points less hostile while their party held the presidency, a pattern robust across specifications, whereas no comparable power effect on schadenfreude survives robustness checks. Cross-national differences that a toxicity-based measure attributes to schadenfreude turn out, under the pleasure-explicit measure, to be differences in hostility. These results show that schadenfreude is a measurable mode of news engagement, distinct from toxicity and patterned by the moral structure of misfortune rather than by country or political power.
\end{abstract}



\maketitle

\section{Introduction}

When a politician is indicted, a company collapses, or a public figure is disgraced, the comment sections of news pages fill with celebration alongside sympathy. Commentators and platform critics have long warned that social media rewards this pleasure at the misfortune of others, turning private amusement into public performance~\cite{papacharissi2014affective, marwick2011tweet, brady2017moral}. The warning is plausible on its face, since digital platforms provide unprecedented visibility into others' setbacks, let users broadcast emotional reactions to wide audiences, and lower the social barriers to voicing publicly what might otherwise remain private or censured~\cite{boyd2014complicated, highfield2016everyday, berger2012viral}. Whether the concern describes a fringe behavior or a routine mode of engagement is unknown, because schadenfreude has scarcely been measured directly and at scale in the environments where it is said to thrive.

Existing evidence leaves three gaps. First, psychological work on schadenfreude rests on laboratory vignettes and self-report~\cite{smith1996envy, vandijk2006grace, feather1999deservingness, leach2015schadenfreude}, which cannot establish how often the emotion is expressed in natural settings, toward which targets, or by which audiences. Second, computational studies of negative engagement measure toxicity, incivility, or outrage~\cite{brady2017moral, brady2021outrage}, constructs that capture hostility but not its defining companion here, pleasure; a comment condemning a wrongdoer and a comment gloating over him are both toxic, yet only the second is schadenfreude. Third, the observational record is confined to single platforms, single countries, and short windows, leaving open whether the patterns attributed to schadenfreude generalize across political and cultural contexts, and whether these patterns have changed over time~\cite{udupa2019extremespeech, pal2015banalities}.

Here we measure expressed schadenfreude in 1{,}064{,}762 public comments on misfortune news posted between 2015 and 2024 by nine major outlets in the United States, the United Kingdom, and India, chosen to span the left--right spectrum within each country. The three countries pair a shared platform and a shared environment of deepening polarization with distinct political cultures and media systems~\cite{iyengar2015affective, newman2021reuters}: a US presidency that changed parties twice in the window, a UK partisan press with a single late transfer of power, and an Indian media environment platformized under one governing party throughout~\cite{sobolewska2020brexitland, udupa2019extremespeech}. This variation lets us separate what is common to misfortune engagement everywhere from what tracks a particular audience's political situation. We define schadenfreude at the comment level as an expression of pleasure, gloating, mockery, or celebration directed at a misfortune, and separate it from hostility without pleasure, sympathy, and neutral reaction. Classification uses an ensemble of three open-weight language models (Qwen3-32B, Llama-3.3-70B, Gemma-3-27B) applied identically to every pair with deterministic decoding and combined by majority vote (Fleiss' $\kappa = 0.66$ across models), a design that any researcher can reproduce exactly and indefinitely; we validate the ensemble against the majority label of three human annotators (accuracy 96.0\%, details in Methods). For the Web community, the study addresses how a platform's affordances shape the public expression of emotion in Web-mediated news engagement.

Post-level metadata, including Facebook's reaction counts and language-model labels for the target, deservedness framing, and event class of each misfortune, let us test where schadenfreude concentrates and how it responds to shifts in political power. We ask four questions: how often audiences express schadenfreude and how the comment measure compares with the platform's reaction icons; how expression differs across outlets and countries, and how those differences compare with hostility; which properties of a misfortune elicit it; and how it responds to shifts in political power.

We find that schadenfreude accounts for 26.1\% of comments on misfortune news, against 31.2\% hostile, 14.4\% sympathetic, and 23.8\% neutral. Comment text reveals more schadenfreude than emoji reactions do. Specifically, the comment-level rate is a third higher than the Haha share of emotional reactions on the same posts (26.1\% against 19.2\%), and the two measures correlate at the post level (Spearman $\rho = 0.43$), indicating that they capture the same disposition. Schadenfreude concentrates where prior theory predicts: on misfortunes framed as deserved (30.5\% against 20.3\% for undeserved), on scandals and defeats rather than deaths, and on public figures rather than ordinary people. In the United States, where the presidency changed parties twice in our window, political power robustly modulates hostility, with the audience of the left-leaning outlet becoming 5.5 percentage points less hostile while its party governed, but has no robust effect on schadenfreude. The distinction between schadenfreude and hostility matters for each of these results: the cross-national patterns that toxicity-based measures attribute to schadenfreude belong, in the contexts that we study here, to hostility.

These findings move schadenfreude from anecdote about platform culture to a measured feature of the news ecosystem. They also carry a lesson for measurement, because once the pleasure component is measured explicitly, schadenfreude is uniform across the countries where toxicity-based measures suggest variation, and that variation is carried by hostility.

\section{Related Work}

\subsection{Schadenfreude in Psychology}
The term schadenfreude combines the German \textit{Schaden} (harm) and \textit{Freude} (joy), and reflection on the emotion runs from Aristotle's \textit{epichairekakia} through Nietzsche's treatment of ressentiment to contemporary moral philosophy~\cite{aristotle1999nicomachean, nietzsche1994genealogy, portmann2000animal, benzeev2000subtle}. Modern psychology organizes schadenfreude around three antecedents. Social-comparison theories hold that the misfortunes of envied or higher-status targets restore threatened self-regard through downward comparison~\cite{smith1996envy}. Deservingness theories locate the pleasure in perceived justice, arising when a wrongdoer suffers a fitting consequence~\cite{feather1999deservingness, vandijk2006grace}. Intergroup theories treat schadenfreude as a group emotion, elicited when an outgroup's setback advances the ingroup's standing in a competitive context such as politics~\cite{tajfel1979socialidentity, leach2015schadenfreude}. Neuroimaging work supports the pleasure component directly, since observing a rival's setback activates reward-related regions including the ventral striatum~\cite{takahashi2009neural}.

Nearly all of this evidence comes from laboratory vignettes and surveys~\cite{webster2024partisan}, and observational studies online have measured schadenfreude through Haha reactions to death coverage~\cite{phillips2024laughing} or in hand-coded samples of a few hundred comments~\cite{cecconi2020schadenfreude}. The three antecedents are also measurable properties of a news post, namely who suffers, whether the suffering is framed as deserved, and whether the target stands in political opposition to the audience, and our design measures each of them at corpus scale.

The experimental literature elicits schadenfreude with material as varied as a rival's exam failure, a celebrity's public humiliation, an athlete's defeat, a company's collapse, and an opposing partisan's scandal~\cite{smith1996envy, vandijk2006grace, leach2015schadenfreude, feather1999deservingness}, and the theories are stated over misfortune in general rather than over any one type. We therefore define misfortune broadly, then label each post's event class (death, scandal, defeat, accident, hardship) and report results within class as well as pooled, so that no single class can drive the aggregate unnoticed.

\subsection{Emotion and Engagement on Platforms}
The ranking and endorsement systems of social platforms promote content that provokes strong reactions~\cite{papacharissi2014affective, bakshy2015facebookdiversity}. Moralized and high-arousal content diffuses farther~\cite{berger2012viral, brady2017moral}, and expressed outrage is reinforced by the engagement it earns~\cite{brady2021outrage}. Platform affordances also convert private feeling into public performance, and ridicule and gloating are central rhetorical devices of networked publics~\cite{marwick2011tweet, phillips2015nice, highfield2016everyday}.

Computational studies of negative engagement, however, measure toxicity, incivility, or outrage, constructs built on insult, harassment, and hate that index aggression toward a target rather than pleasure at its misfortune. The two behaviors have different theoretical antecedents (anger versus reward) and different elicitors, so we measure both in the same corpus and compare how they pattern.

\subsection{News Audiences and Polarization}
Audiences sort into congenial outlets and interpret information through partisan lenses~\cite{stroud2011niche}, and affective polarization has made outparty misfortune an occasion for gratification across democracies~\cite{iyengar2015affective}. These dynamics are conditioned by national media systems. In the United Kingdom, the Brexit era deepened identity-driven cleavages within a long-standing partisan press~\cite{sobolewska2020brexitland}. In India, the platformization of news has entangled digital media with partisan and communal mobilization, and ``extreme speech'' scholarship documents the normalization of derision online~\cite{pal2015banalities, udupa2019extremespeech, newman2021reuters}. In the United States, Facebook is a major gateway to political news~\cite{bakshy2015facebookdiversity}. No prior study compares expressed schadenfreude across such systems, over a period long enough to contain multiple transfers of political power.

\subsection{Research Questions}\label{sec:rqs}
Against this background we ask four questions. \textbf{RQ1:} How frequently do audiences express schadenfreude in response to news of others' misfortunes, and how does the comment-based measure compare with reaction icons on the same posts? \textbf{RQ2:} How does expressed schadenfreude differ across outlet leanings and national contexts, and how do those differences compare with hostility? \textbf{RQ3:} Which properties of a misfortune, its deservedness framing, event class, target, and topic, most reliably elicit schadenfreude? \textbf{RQ4:} How does expression respond to shifts in political power?

Each question concerns schadenfreude \emph{as expressed in a public comment section}, a setting with an audience, a moderation regime, and a permanent record. People also feel and voice schadenfreude in private, where none of those conditions hold, and our data do not estimate how often they do so. The public comment section therefore understates how often the emotion is felt, while measuring directly how often it is expressed.

\section{Data and Methods}

\subsection{Data Collection}
We collected posts, public comments, and reaction counts from the verified Facebook pages of nine news outlets: NBC News (left), CNBC (center), and Fox News (right) in the United States; The Guardian (left), Reuters (center), and The Telegraph (right) in the United Kingdom; and Hindustan Times (left), Indian Express (center), and OpIndia (right) in India. Outlet leanings follow published media-bias ratings (AllSides; Media Bias/Fact Check). Within each country we selected one outlet per leaning, choosing at each position the outlet whose Facebook page had the largest following at the time of data collection. We additionally collected a tenth outlet, Metro (UK; rated Left-Center), which enters only as a robustness check in Appendix~\ref{app:metro} because it duplicates an ideological cell already covered by The Guardian.
For each outlet we collected up to 1{,}000 posts per month from 2015 through 2024 with their text, timestamps, public comments, and reaction counts, by automated collection from each page's public timeline. Within each month, posts were collected in reverse chronological order, so a month that exceeded the cap would have lost its earliest posts. The cap applied to all of a page's posts, of which misfortune posts are a subset; in the analyzed corpus the busiest dated outlet-month contains 302 misfortune posts and the median 17. The coverage reported in Appendix Table~\ref{tab:transparency} therefore reflects each outlet's posting and comment-retention history and, in any month in which a page published more than 1{,}000 posts, the cap. We restrict analysis to publicly visible content with identifying information removed. The main corpus contains 20{,}661 misfortune posts and 1{,}064{,}762 associated comments from the nine outlets; Metro adds 3{,}687 posts and 99{,}244 comments used only in the appendix. Appendix Table~\ref{tab:transparency} reports posts and comments per outlet and year, including known coverage gaps.

\subsection{Identifying Misfortune Posts}
Three trained annotators labeled a random sample of 500 collected posts as describing a misfortune, defined as harm, scandal, failure, or loss affecting a person, group, or organization, or not; inter-annotator agreement was high ($\kappa = 0.81$). The annotations refined and validated a language-model classifier (GPT-4o Mini) that we then applied to all collected posts (accuracy 94.8\%, precision 88.9\%, recall 87.3\%, F1 88.1\% against the held-out human labels). This corpus-selection step is the one stage of the pipeline that uses a commercial model; the post-level labels and every comment label, which carry the results, use open weights. All subsequent analysis concerns the posts classified as misfortune.

The retained posts span all five misfortune subclasses in quantity, including the ones prior work identifies as most schadenfreude-relevant (scandal $n = 8{,}125$, defeat $n = 654$, accident $n = 3{,}511$, hardship $n = 4{,}452$, death $n = 3{,}888$). Appendix~\ref{app:measurement} explains why we classify rather than retrieve by keyword.

\subsection{Measuring Schadenfreude}
A comment can insult without gloating and can gloat without insulting, so toxicity and schadenfreude are different constructs. The computational literature on negative engagement operationalizes constructs of this kind with toxicity-style classifiers (insults, harassment, hate), an operationalization that leaves the defining component, pleasure, unmeasured. To quantify what that measurement choice does, we also label the full corpus with a fine-tuned toxicity-style classifier (toxic, sympathetic, neutral, unknown) and show that the two constructs diverge empirically (Section~\ref{sec:results}). We therefore classify every post--comment pair into five categories: \textbf{SCH} (schadenfreude: pleasure, amusement, celebration, gloating, or mockery at the misfortune), \textbf{HOS} (hostility without pleasure: condemnation, blame, or insult with no expressed enjoyment), \textbf{SYM} (sympathy), \textbf{NEU} (neutral), and \textbf{UNK} (unclassifiable). The full annotation scheme, including rules for sarcasm and mixed cases, appears in Appendix~\ref{app:scheme}.

Every collected post--comment pair (1{,}164{,}108 including the Metro robustness set) was classified independently by three open-weight instruction-tuned models from three developers, Qwen3-32B, Llama-3.3-70B-Instruct, and Gemma-3-27B-IT, using an identical prompt (Appendix~\ref{app:prompt}) at temperature 0. The final label is the majority vote; pairs with no majority are labeled UNK. Inter-model agreement is substantial (Fleiss' $\kappa = 0.66$; pairwise agreement 73--77\%).
We use open-weight models for reproducibility and reuse. Pinned model versions and deterministic decoding make the classification exactly reproducible by anyone, indefinitely, whereas a measurement built on a commercial endpoint stops being reproducible once that endpoint is updated or retired. The same property makes the pipeline reusable: the weights, the prompt, and the vote are the whole method, so a later study, or a platform-integrity team, can rerun it on new data or extend it to another language without our cooperation and without an API budget. Off-the-shelf toxicity and emotion classifiers cannot do this labeling, for reasons set out in Appendix~\ref{app:measurement}; instead of discarding that comparison we make it an object of study, labeling the same corpus with a toxicity-style classifier and reporting where the two measures disagree (Section 4.6). Against a stratified gold set of 200 pairs labeled by three human annotators, the ensemble reaches accuracy 96.0\% (95\% CI 92.3--98.0\%) and macro-F1 0.93; per-model and per-class results, and a comparison against the toxicity operationalization, appear in Appendix~\ref{app:validation}.
We correct all reported prevalence estimates for classifier error using confusion-matrix inversion on the gold set, extending the two-class estimator of \citet{rogan1978estimating} to five classes, and report both raw and adjusted figures.

The gold set consists of 200 post--comment pairs sampled with stratification by outlet (20 per outlet) and spread across years; the sample contains all four of the toxicity classifier's labels (67 toxic, 78 neutral, 29 sympathetic, 26 unknown). Three annotators labeled the set independently, without access to the model labels, under written guidelines with hard-case rules (Appendix~\ref{app:scheme}), and one annotator left a single pair blank. The gold label is the majority label; the five pairs on which all three annotators disagreed were adjudicated by one annotator with the model labels hidden. Agreement among the annotators was substantial (Fleiss' $\kappa = 0.76$ over the 199 pairs with three labels; pairwise Cohen's $\kappa$ 0.70--0.81). Because the same 200 pairs carry the toxicity classifier's labels, they serve two purposes, validating the ensemble and quantifying, at the human level, how often toxicity and schadenfreude judgments diverge.

\subsection{Post-Level Properties}
Each unique post was labeled by a language model (Qwen3-32B, temperature 0) for the target of the misfortune (ordinary person, public figure, organization, group), its deservedness framing (deserved, undeserved, none), its event class (death, scandal, defeat, accident, hardship), and its topic. The eight topic domains follow the antecedents identified in prior work: political, religious, and institutional and moral or ideological misfortunes capture moral transgression and ideological conflict; economic and corporate and social and group-level misfortunes reflect competitive and hierarchical dynamics; celebrity, personal, and sports misfortunes represent lower-stakes social comparison; and natural and environmental events serve as a contrast category of non-agentic harm that typically elicits empathy. Parse failures were below 0.2\% for every field. These labels are validated on a separate post-level gold set of 180 posts, 20 per outlet and spread across all ten years, which two annotators labeled in full and the third labeled on a subset of 59 posts, for all four fields under the guidelines of Appendix~\ref{app:scheme}. Against the majority label on the 59 triple-annotated posts, Qwen's accuracy is 100\% for topic, 98.3\% for target, and 94.9\% for both deservedness and event class (Appendix~\ref{app:validation}).
Reaction counts (Love, Care, Haha, Wow, Sad, Angry) are observed for 84\% of posts; Fox News posts, collected in a separate 2024 scrape, lack reaction metadata and post dates.

\subsection{Statistical Approach}
The unit of analysis is the post: we compute the share of a post's comments in each category and aggregate or model at the post level, weighting variants by comment count. Inference uses heteroskedasticity-robust (HC1) standard errors. For the political-power analysis we estimate, for post $p$ from outlet $o$ at time $t$,
\begin{equation}
y_{p} = \beta_0 + \beta_1 \mathrm{Right}_o + \beta_2 \mathrm{InPower}_{ot} + \beta_3 \mathrm{Right}_o \times \mathrm{InPower}_{ot} + \gamma_{\mathrm{topic}(p)} + \varepsilon_p,
\end{equation}
where $y_p$ is the post's schadenfreude (or hostility) share, $\mathrm{InPower}_{ot}$ indicates that the party aligned with outlet $o$'s audience holds the presidency at $t$, coded to inauguration day where post dates exist, and $\gamma$ are topic fixed effects; we report ordinary and comment-weighted least squares. All reported significance tests are corrected for multiple comparisons with Benjamini--Hochberg false-discovery-rate control across the full grid of reported cells. Analyses labeled exploratory were not part of the pre-specified grid. Because each country--leaning cell contains a single outlet, we present per-outlet results throughout and phrase claims as statements about the audiences of these outlets rather than about national or ideological populations.

The study analyzes only public content, reports only aggregates, and involves no interaction with users; ethical considerations are detailed in Appendix~\ref{app:ethics}.

\section{Results}\label{sec:results}

\subsection{Reactions to misfortune news}
\begin{figure*}[t]
\centering
\includegraphics[width=0.9\linewidth]{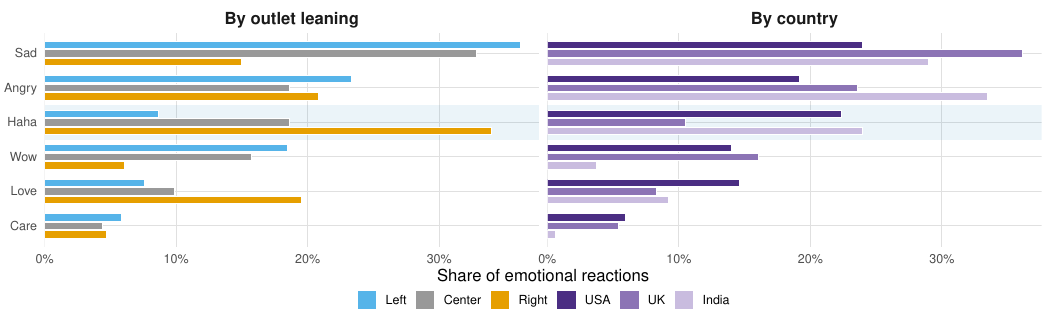}
\caption{Amusement at misfortune is visible in emoji reactions and varies with audience; Haha is the most common emotional reaction among right-leaning audiences. Bars give the composition of emotional reactions on misfortune posts, by outlet leaning (left panel) and country (right panel); the shaded row marks Haha.}
\label{fig:reactions}
\end{figure*}

We first examine the coarsest layer of engagement, Facebook's emoji reactions, on the 17{,}395 misfortune posts with reaction metadata. Sad (27.9\%) and Angry (21.9\%) dominate, so audiences primarily process misfortune through empathic sorrow and moral outrage. Haha, the reaction closest to expressed pleasure, still accounts for 19.2\%, ahead of Wow (13.5\%) and Love (12.3\%). The composition shifts with outlet leaning: among right-leaning outlets Haha is the single most common emotional reaction (34.0\%), against 8.7\% for left-leaning audiences, whose modal reaction is Sad (36.1\%). Indian (23.9\%) and US (22.3\%) audiences react with Haha at roughly twice the UK rate (10.5\%); Figure~\ref{fig:reactions} gives the full composition by leaning and country. Indian audiences are also the angriest (33.5\% Angry against 23.6\% in the UK and 19.2\% in the US), the same ordering the comment-level hostility rates show in Section 4.2. Reaction icons, therefore, already show that amusement at misfortune is common and socially patterned.

 They are also, as the next sections show, a substantial undercount.

\subsection{Schadenfreude is a quarter of the conversation}
We next ask how much of the response to misfortune news is schadenfreude rather than some other reaction. Across 1{,}064{,}762 classified post--comment pairs, 26.1\% of comments express schadenfreude, 31.2\% are hostile without pleasure, 14.4\% are sympathetic, 23.8\% are neutral, and 4.4\% could not be classified by majority vote (Table~\ref{tab:prevalence}). Appendix~\ref{app:examples} gives illustrative comments for each category.

\begin{table}[t]
\centering\small
\caption{Comment-level label distribution (\%), overall and by country. Raw classifier output; error-adjusted estimates follow the gold-set validation (Appendix~\ref{app:validation}).}
\label{tab:prevalence}
\begin{tabular}{lrrrrr}
\toprule
 & SCH & HOS & SYM & NEU & UNK \\
\midrule
All comments & 26.1 & 31.2 & 14.4 & 23.8 & 4.4 \\
United States ($n$=679{,}041) & 26.2 & 31.8 & 14.7 & 23.3 & 4.0 \\
United Kingdom ($n$=275{,}724) & 26.1 & 27.5 & 14.3 & 27.0 & 5.1 \\
India ($n$=109{,}997) & 25.8 & 37.5 & 12.9 & 18.9 & 4.8 \\
\bottomrule
\end{tabular}
\end{table}

\begin{figure}[t]
\centering
\includegraphics[width=\linewidth]{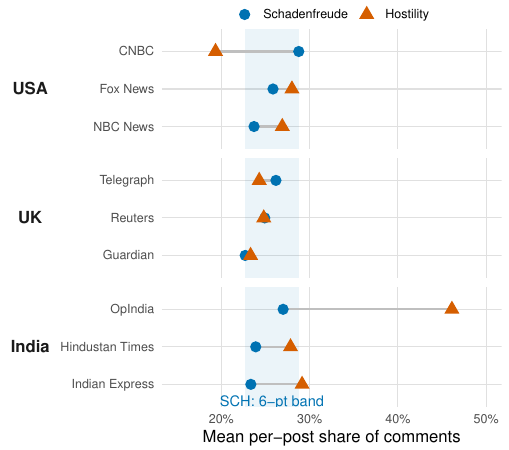}
\caption{Schadenfreude is nearly uniform across outlets, while hostility varies widely. Circles give the mean per-post share of comments expressing schadenfreude (SCH) and triangles hostility without pleasure (HOS), by outlet and country. The shaded band marks the six-point range that contains every outlet's schadenfreude rate; hostility varies by 27 points and produces the cross-outlet differences that toxicity-based measures attribute to schadenfreude.}
\label{fig:outlets}
\end{figure}

 Averaging within posts first, the mean schadenfreude rate per post ranges from 22.7\% (The Guardian) to 28.8\% (CNBC) across the nine outlets (Figure~\ref{fig:outlets}), a band of six percentage points spanning three countries, three ideological leanings, and nine distinct moderation regimes.

\begin{figure*}[t]
\centering
\includegraphics[width=0.9\linewidth]{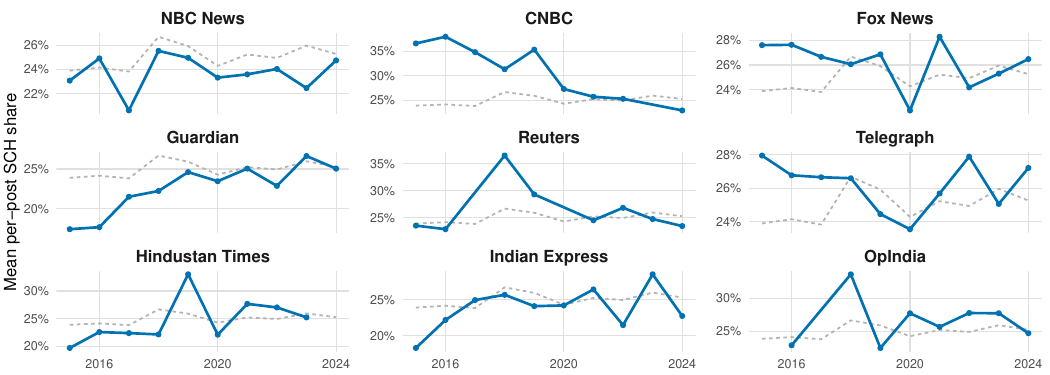}
\caption{No outlet's schadenfreude rate trends over the decade, and the two that move (CNBC down, The Guardian up) converge on the common level. Each panel is one outlet's yearly mean per-post schadenfreude share (blue), against the pooled yearly mean across all nine outlets (dashed grey). Each panel's vertical axis is scaled to that outlet's own range, so slopes are comparable within panels but not across them; outlet-years with fewer than 30 posts are suppressed.}
\label{fig:time}
\end{figure*}

The rate is equally stable over time, with yearly averages between 23.8\% and 26.7\% across the decade and no sustained trend in any country (Figure~\ref{fig:time}), although individual outlets move within that stability. CNBC's audience declined from 36.6\% in 2015 to 22.9\% in 2024, converging on the other US outlets; The Guardian's rose from 17.5\% to 25.0\%, converging from below; and NBC and Fox News audiences stayed within four points of their decade means throughout (Figure~\ref{fig:time}). Overall, the outlets regress toward a common level, with no platform-wide rise or fall in expressed pleasure at misfortune.

These estimates are raw classifier outputs. Correcting them for classifier error with the human gold set changes no share by more than 1.8 points (schadenfreude 25.1\%, 95\% CI 22.3--27.2\%; hostility 32.7\%) and leaves the three countries' schadenfreude shares within 0.5 points of each other (Appendix~\ref{app:validation}).

The uniformity of schadenfreude contrasts with the behavior of hostility. Hostile-without-pleasure rates vary from 19.3\% (CNBC) to 46.1\% (OpIndia), and the cross-national ordering that toxicity-based measures attribute to schadenfreude is in fact an ordering of hostility: India's per-post hostility rate (36.4\%) exceeds that of the United States (26.9\%) and the United Kingdom (24.1\%), while the three countries' schadenfreude rates sit within one percentage point of each other (25.1\%, 25.3\%, 24.5\%). India's elevated hostility is concentrated in OpIndia; excluding it lowers the Indian schadenfreude average from 25.1\% to 23.6\%. Audiences of right-leaning outlets show both more schadenfreude (26.3\% versus 23.4\% for left-leaning) and substantially more hostility (32.5\% versus 26.0\%).

\subsection{Comments reveal what reactions conceal}
Facebook's reaction buttons offer a low-cost, publicly visible channel for the same disposition, so we test whether our text-based measure and the platform's ``Haha'' reaction agree. Among the 10{,}970 posts with at least five comments and observed reactions, the per-post schadenfreude rate correlates with the Haha share of emotional reactions at Spearman $\rho = 0.43$ (India 0.49, United Kingdom 0.42, United States 0.40), whereas hostility shows no such association ($\rho = 0.07$). The two channels differ in level as well: schadenfreude appears in 26.1\% of comments but Haha constitutes 19.2\% of emotional reactions, and the latter figure is insensitive to the post-inclusion threshold (19.2\% across minimum-reaction cutoffs of 0 to 100). Reaction counts therefore understate expressed pleasure relative to comment text, although the two shares are computed over different populations (commenters and reactors), so we treat the difference in level as indicative and rest the validation on the post-level correlation.

\subsection{Deservingness, targets, and events}
\begin{figure*}[t]
\centering
\includegraphics[width=0.9\linewidth]{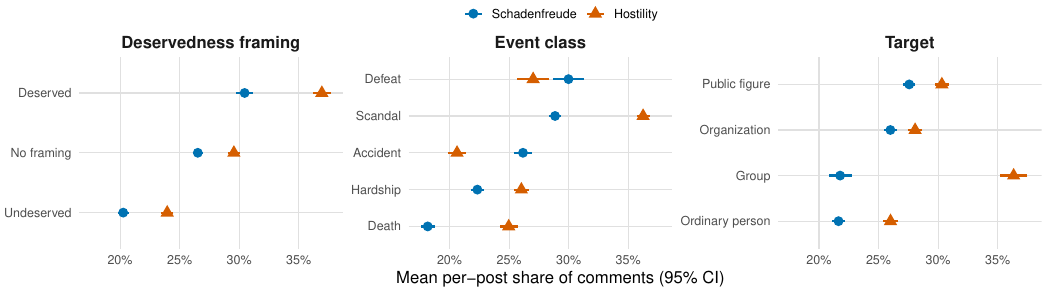}
\caption{Schadenfreude varies with the deservedness framing, event class, and target of the misfortune, and hostility follows a different pattern. Points give the mean per-post share of schadenfreude (circles) and hostility (triangles) with 95\% confidence intervals, by the post's deservedness framing, event class, and target type; panels share the horizontal scale. Schadenfreude peaks for deserved-framed misfortunes, defeats and scandals, and public figures; hostility peaks for scandals and group targets, and stays high even for deaths.}
\label{fig:moderators}
\end{figure*}

Psychological accounts predict that schadenfreude concentrates on misfortunes that can be read as deserved, on high-status targets, and on reputational rather than mortal harms, and our data support all three predictions (Figure~\ref{fig:moderators}).

Posts that frame the misfortune as the target's own doing draw schadenfreude in 30.5\% of comments against 20.3\% for posts framing the target as an innocent victim, a gap of $+3.6$ percentage points ($[+2.7, +4.5]$) after conditioning on event class and target type. Relative to deaths, the schadenfreude rate is 8--11 points higher for scandals ($+8.6$ $[+7.7, +9.4]$), defeats ($+10.6$ $[+9.1, +12.0]$), and accidents ($+8.4$ $[+7.4, +9.4]$). Public figures attract 2.9 points more schadenfreude than ordinary people ($[+2.1, +3.7]$), and organizations 1.7 points more. Hostility follows a different pattern, highest for scandals and for group targets, and, unlike schadenfreude, it exceeds 24\% even for deaths.

Because event class is by far the strongest moderator, it is also the most plausible confound for the cross-national uniformity of Section 4.2: three countries could reach the same average by mixing different events. Within every event class, however, the three countries land within three and a half points of each other, and usually within one (deaths 18.1--18.4\%, hardships 22.1--22.5\%, scandals 28.3--29.4\%, accidents 25.5--27.3\%, defeats 27.4--30.9\%, the last on the smallest cells), so audiences in the three countries respond to the same kind of misfortune at the same rate and the uniformity is not an artifact of event composition. All coefficients survive Benjamini--Hochberg correction across the full reported grid.

\begin{figure}[t]
\centering
\includegraphics[width=\linewidth]{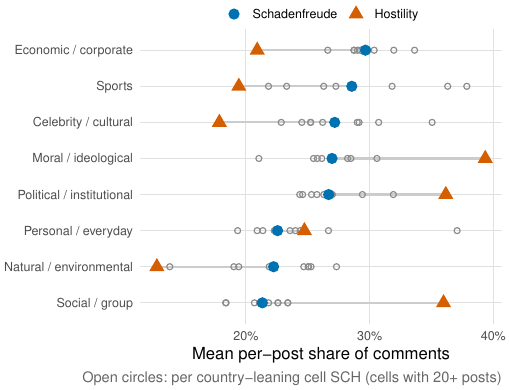}
\caption{The topics that draw the most hostility differ from those that draw the most schadenfreude. Filled circles give each topic's pooled mean per-post schadenfreude share and triangles its hostility share; open circles give the schadenfreude share of each country--leaning cell separately (cells with fewer than 20 posts omitted). Topic labels come from the same open-weight pipeline as the comment labels.}
\label{fig:topics}
\end{figure}

The topic domains tell the same story at a coarser grain (Figure~\ref{fig:topics}). Schadenfreude peaks for economic and corporate (29.7\%), sports (28.6\%), and celebrity (27.2\%) misfortunes, the domains where targets are elites or rivals and stakes fall short of tragedy, and bottoms out for social and group-level (21.3\%), natural and environmental (22.2\%), and personal (22.6\%) misfortune. Hostility concentrates elsewhere entirely: moral and ideological (39.4\%), political and institutional (36.2\%), and social and group-level (36.0\%) misfortunes, the morally and politically charged domains, draw the most anger while sitting mid-range for pleasure. The topics that draw the most hostility therefore differ from those that draw the most schadenfreude, which shows the SCH--HOS distinction at the level of subject matter. Within each topic the spread across the nine country--leaning cells is modest (Figure~\ref{fig:topics}, open circles), consistent with the cross-outlet uniformity of Section 4.2.

\subsection{Political power and its loss}
The United States is the one country in our window with two transfers of presidential power, so we estimate how engagement responds when the party aligned with an outlet's audience gains or loses the White House. We model the per-post schadenfreude rate as a function of audience alignment with the governing party, coded to the day of inauguration, with topic fixed effects and HC1 errors, on 7{,}791 post cells from NBC and Fox News (Figure~\ref{fig:power}; full estimates in Appendix Table~\ref{tab:regression}).
For hostility the pattern is consistent across every specification: comments on the left-leaning outlet become substantially less hostile when Democrats hold the presidency ($-5.5$ points unweighted $[-6.9, -4.1]$; $-6.3$ weighted $[-9.2, -3.4]$), and the interaction for the right-leaning outlet cancels the effect almost exactly ($+6.3$ $[+4.5, +8.1]$), leaving Fox audiences' hostility unmoved by who governs. This opposition-hostility pattern survives dropping the transition years 2017 and 2021 ($-3.8$ $[-5.5, -2.1]$) and comment weighting. For schadenfreude, by contrast, the power effects are small and fragile: the in-power main effect is $+1.1$ points ($[-0.1, +2.4]$) and the right-leaning interaction $-1.8$ points ($[-3.5, -0.2]$), but both fall to null when transition years are excluded (Table~\ref{tab:regression}). In Figure~\ref{fig:power}, every hostility interval sits well clear of zero, and every schadenfreude interval touches or nearly touches it. We conclude that political power reliably modulates the hostility of opposition audiences, and we find no robust evidence that it modulates schadenfreude.

\begin{figure*}[t]
\centering
\includegraphics[width=0.9\linewidth]{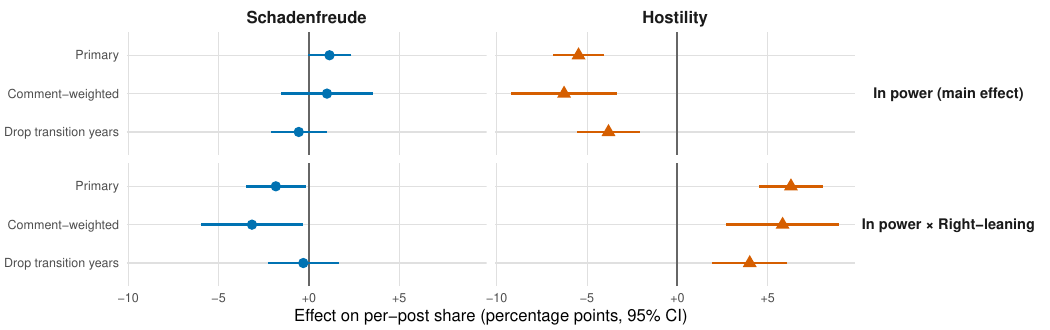}
\caption{Political power shifts the hostility of opposition audiences but has no robust effect on schadenfreude. Each estimate is the effect of the audience's party holding the presidency on the per-post share of comments in that category, with 95\% confidence intervals, for the three US specifications (primary, comment-weighted, transition years dropped). Circles are schadenfreude, triangles hostility; the vertical line is the null. Hostility effects are large and consistently signed; schadenfreude effects are small and do not survive dropping the transition years.}
\label{fig:power}
\end{figure*}

Extending the analysis beyond the United States illustrates why the design needs power to actually change hands. The United Kingdom's single transfer (Labour taking office in July 2024) leaves six months of post-change data against nine and a half years before it; on 3{,}874 dated post cells from The Guardian and The Telegraph, Guardian audiences' schadenfreude moved from 22.8\% to 24.0\% after the change and Telegraph audiences' from 28.6\% in their brief opposition spell to 26.1\% in the long governing one, but no coefficient is statistically reliable ($+1.2$ points $[-2.4, +4.8]$ for the in-power main effect), and we present the country as suggestive only. India experienced no transfer of national power in our window, with the same party governing from 2014 through 2024, so a within-audience power effect is not identifiable there at all; Indian audiences contribute to every other analysis but not to this one.

\subsection{Schadenfreude is not toxicity}
The preceding results repeatedly separate schadenfreude from hostility; we close by quantifying the gap against the toxicity-style operationalization directly. Of the comments the fine-tuned toxicity classifier labels toxic, 58.1\% are HOS and 34.9\% SCH under the pleasure-explicit scheme, so a toxicity measure of schadenfreude counts five hostile comments for every three gleeful ones. The error runs in the other direction as well: 28.9\% of the comments the toxicity classifier calls neutral are SCH under the pleasure-explicit scheme, largely bare laughing emojis, ``karma'' remarks, and mockery that contains no insult. At the post level the two constructs are negatively correlated (Pearson $r = -0.28$ between a post's SCH and HOS shares), so posts that draw more pleasure tend to draw less anger. Together with the divergent Haha correlations (Section 4.3) and the divergent political responses (Section 4.5), this places schadenfreude and hostility on separate empirical footings.

\subsection{Robustness}\label{sec:robustness}
Three features of the data could mechanically produce our patterns: a small set of prolific commenters, the particular outlets chosen, and coding decisions in the political-power analysis. We find no evidence that any of the three does. Excluding each outlet's top 1\% of commenters moves no outlet's schadenfreude rate by more than 0.8 points; dropping OpIndia lowers India's average from 25.1\% to 23.6\%, still within two points of the other countries; and the hostility power results hold under all three in-power codings, while the schadenfreude interaction, which does not, is flagged as non-robust above. Appendix~\ref{app:robustness} gives the full checks.

\section{Discussion}

Platform critics have warned for a decade that social media turns the misfortunes of others into entertainment. Our measurements support this concern, since pleasure at misfortune is a routine register of news engagement, roughly a quarter of all comments on misfortune news across nine outlets, three countries, and ten years. However, the behaviors that critics group together under toxicity separate in our data, and three findings organize what follows.

First, schadenfreude and hostility are empirically distinct behaviors. They correlate differently with the platform's own pleasure signal (Spearman $\rho = 0.43$ against 0.07 with Haha reactions), concentrate on different events and targets, and respond differently to political power. Toxicity-based measures mix the two and inherit the patterns of whichever component dominates the corpus at hand. Under a toxicity operationalization, for example, Indian audiences appear far more schadenfreude-prone than British or American ones, whereas with pleasure measured explicitly the three countries are indistinguishable and India's distinctiveness is anger, concentrated in a single outlet (OpIndia).

Second, schadenfreude follows deservingness. Misfortunes framed as self-inflicted, scandals and defeats rather than deaths, and public figures rather than ordinary people all elicit more of it, matching laboratory accounts in which deserved suffering is the core elicitor and mortal harm suppresses pleasure. The antecedents identified in the laboratory therefore also organize a decade of naturally occurring behavior.

Third, political power moves hostility but not schadenfreude. Audiences of the left-leaning US outlet became 5.5 percentage points less hostile while their party governed, under every specification we examined, whereas the hostility of the right-leaning outlet's audience did not respond to power. Schadenfreude, in the same models on the same posts, does not move reliably with power.

It would be a mistake to read these results as claims about national cultures or ideological populations. Each country--leaning cell contains a single outlet, so every comparison across leanings or countries is a comparison of particular audiences, and the English-language Indian outlets in our sample reach an anglophone minority rather than Indian audiences at large.

Two mechanisms, likely compounding, could explain why schadenfreude is uniform where hostility varies. Schadenfreude tracks properties of the misfortune itself (its deservingness framing, event class, and target status), which news of scandal and failure supplies at similar rates everywhere, whereas hostility tracks the audience and its political situation. Any audience can also construct a story in which the target had it coming, while sustained anger requires a grievance, and grievances are unevenly distributed by political position. Because our data are audience aggregates, neither mechanism implies that schadenfreude is fixed at the individual level, but the uniformity is consistent with pleasure at deserved misfortune being a stable human response that platforms make visible, with hostility as the channel for political grievance.

For affective polarization, the results separate two emotional components that respond to different conditions. The pleasure component of outparty misfortune appears wherever deservingness can be constructed, on both sides of the spectrum and in all three countries, while the anger component carries the political signal. Interventions that target ``toxicity'' as a single construct will therefore hit two behaviors with different drivers, and may misread a politically driven surge in anger as a rise in cruelty, or the reverse. The measurement pipeline is also reusable, since three open-weight models, a fixed prompt, and majority vote reproduce our labels exactly at zero marginal cost, on hardware a platform-integrity or civil-society team can afford.

\subsection{Limitations and Future Work}
Several limitations follow from the design. First, our labels capture expressed reaction, not felt emotion; sarcasm and performative gloating are classified as what they display (accuracy against human judgment 96.0\%, Appendix~\ref{app:validation}). Second, error compounds along the pipeline: a comment enters our estimates only if its post was correctly classified as a misfortune (94.8\% accurate) and its own label is correct, and the confusion-matrix correction we apply operates on the comment stage only, so prevalence levels are more uncertain than the comparisons between them, which share the same pipeline and largely difference the error out. Third, the design is observational, and outlet audiences self-select, so power effects are within-audience changes over time, not causal effects of power on individuals. Fourth, the study observes one platform, whose reaction affordances, comment ranking, and moderation regime are part of what we measure, so levels of expressed schadenfreude elsewhere need not match ours, although the mechanism claims are more likely to generalize than the point estimates. Fifth, comment availability reflects each outlet's moderation and Facebook's retention, with documented gaps (Appendix~\ref{app:transparency}). Automated accounts are a related concern that we can limit but not exclude: we have no bot labels, but the heavy-user check (Section~\ref{sec:robustness}) removes the most prolific commenters, where coordinated posting would concentrate, without moving any rate by more than 0.8 points. Finally, our gold-standard validation rests on 200 majority-labeled pairs, 59 triple-annotated posts, and three independent annotators; a larger, multilingual gold set covering Hindi--English code-switching is the clearest next step for the Indian corpus.

\bibliographystyle{ACM-Reference-Format}
\bibliography{references}

\appendix

\section{Annotation Scheme}\label{app:scheme}
Every post in the corpus describes a misfortune; the scheme classifies how the commenter reacts to it. \textbf{SCH} (schadenfreude): the commenter expresses pleasure, amusement, satisfaction, celebration, gloating, or mockery at the misfortune or at those affected, including ``karma'' and ``serves them right'' remarks, laughing emojis at a downfall, sarcastic sympathy, and jokes at the victim's expense. \textbf{HOS} (hostile without pleasure): anger, condemnation, blame, insult, or hate with no enjoyment of the misfortune, including moral outrage, demands for punishment, and flat attributions of fault. \textbf{SYM} (sympathetic): empathy, support, kindness, or condolence toward those affected. \textbf{NEU} (neutral): factual, informational, indifferent, or off-topic comments, tags, questions, and reactions to headline wording rather than to the misfortune. \textbf{UNK} (unknown): too ambiguous or unintelligible to classify. Two adjudication rules resolve the hard boundary: any real enjoyment of the misfortune makes a comment SCH even when it also insults, and ``deserved it'' is SCH when delivered with glee but HOS when delivered flat. A bare laughing emoji on a misfortune post is SCH.

\section{Robustness Checks}\label{app:robustness}
\textbf{Heavy users.} The top 1\% of commenters contribute 7--15\% of comments per outlet, and comment sections select for the most engaged users generally. Excluding each outlet's top 1\% moves no outlet's schadenfreude rate by more than 0.8 percentage points, and the moderator orderings are unchanged. Fox News comments carry no user identifiers, so this check covers the other eight outlets.

\textbf{Outlet composition.} India's schadenfreude average moves from 25.1\% to 23.6\% without OpIndia, remaining within two points of the other countries; adding the tenth outlet, Metro, to the United Kingdom leaves the UK average essentially unchanged (24.5\% against 24.5\% without it; Appendix~\ref{app:metro}). Because every country--leaning cell contains a single outlet, we report per-outlet values throughout (Figure~\ref{fig:outlets}) rather than treating any aggregate as a national estimate.

\textbf{Coding decisions.} Power-analysis results are reported under day-of-inauguration coding, majority-of-year coding, and transition-year exclusion (Table~\ref{tab:regression}); the hostility findings hold under all three, and the schadenfreude interaction does not survive transition-year exclusion. The reaction-based figures are insensitive to the post-inclusion threshold (Haha share 19.2\% across minimum-reaction cutoffs from 0 to 100).

\section{Example Comments}\label{app:examples}
\begin{table}[h]
\centering\small
\caption{Illustrative comments for each category, paraphrased from corpus examples to preserve commenter anonymity. Post context: a news item reporting a public figure's fraud conviction.}
\label{tab:examples}
\begin{tabular}{lp{5.6cm}}
\toprule
SCH & ``Karma finally caught up with him'' [with laughing emoji]; ``couldn't have happened to a nicer guy'' \\
HOS & ``Lock him up and throw away the key, he's a disgrace'' \\
SYM & ``I feel for his family, they didn't ask for any of this'' \\
NEU & ``What does this mean for the appeal?'' \\
UNK & ``ok then'' \\
\bottomrule
\end{tabular}
\end{table}

\section{Measurement Choices}\label{app:measurement}
\textbf{Why classify misfortune rather than retrieve it.} Retrieving posts by subclass keywords (\textit{died}, \textit{arrested}, \textit{convicted}, \textit{layoffs}, \textit{defeat}) would be more auditable, but it miscounts where this study is most sensitive. Such a filter cannot separate a report of a conviction from a report of an acquittal, and it misses the paraphrase and euphemism that headline writers favor. Its recall would also vary with each outlet's house style, confounding the cross-outlet comparison that is our main object. What a keyword view does supply is a coverage check, which the classified corpus passes (subclass counts in Section~\ref{sec:results}).

\textbf{Why not an off-the-shelf toxicity or emotion classifier.} No released toxicity model carries a class for pleasure at another's misfortune, and emotion models return a valence for the comment rather than a stance toward the event the post describes. A gloating one-line ``karma'' comment therefore reads as low-toxicity and mildly positive in both families, which is precisely the case our measure must catch. Section 4.6 reports what a toxicity-style classifier does to the schadenfreude estimate when used as a proxy.

\section{Classification Prompt}\label{app:prompt}
The three ensemble models received the following prompt for every post--comment pair, at temperature 0, with the post truncated to 1{,}200 characters and the comment to 1{,}200 characters:

\begin{quote}\small\ttfamily
You are annotating Facebook comments on news posts. Every post describes a misfortune (harm, scandal, failure, or loss affecting a person, group, or organization). Classify how the COMMENTER reacts to that misfortune. Attend to sarcasm, irony, and emojis.\\[2pt]
Labels:\\
SCH --- Schadenfreude: the commenter expresses pleasure, amusement, satisfaction, celebration, gloating, or mockery at the misfortune or at those affected (e.g., "karma", "serves them right", laughing emojis at someone's downfall, sarcastic sympathy, jokes at the victim's expense).\\
HOS --- Hostile without pleasure: anger, condemnation, blame, insults, or hate with no enjoyment of the misfortune (moral outrage, demands for punishment, flat "he brought this on himself").\\
SYM --- Sympathetic: empathy, support, kindness, or condolence toward those affected.\\
NEU --- Neutral: factual, informational, indifferent, off-topic, tags, questions, or reactions to the headline wording rather than the misfortune.\\
UNK --- Unknown: genuinely too ambiguous or unintelligible to classify.\\[2pt]
Rules: if any real enjoyment of the misfortune is present, SCH wins over HOS. A bare laughing emoji on a misfortune post is SCH. "Deserved it" with glee is SCH; without glee it is HOS.\\[2pt]
Post: "\{post\}"\\
Comment: "\{comment\}"\\[2pt]
Answer with exactly one label: SCH, HOS, SYM, NEU, or UNK.
\end{quote}

\section{Validation}\label{app:validation}
Inter-model agreement over the full corpus: Qwen--Llama 73.4\%, Qwen--Gemma 76.6\%, Llama--Gemma 74.6\%; Fleiss' $\kappa = 0.664$.

Three annotators labeled the 200 gold pairs independently, and one annotator left a single pair blank. Agreement among the annotators was substantial (Fleiss' $\kappa = 0.764$ over the 199 pairs with three labels; all three agreed on 74.9\% of pairs; pairwise Cohen's $\kappa$ 0.704--0.813 and raw agreement 77.9--85.9\%). The majority label decided 195 pairs, and the remaining five, on which all three annotators disagreed, were adjudicated by one annotator with the model labels hidden. Against this gold label the ensemble reaches 96.0\% accuracy (Wilson 95\% CI 92.3--98.0\%) and macro-F1 0.929, above each of the three models alone (84.0--87.5\%; Table~\ref{tab:gold}), so the vote corrects errors that the models do not share. Per-class F1 for the ensemble is 0.98 for hostility, 0.97 for neutral, 0.96 for schadenfreude, 0.96 for sympathy, and 0.78 for unknown.

The ensemble and the majority label disagree on eight pairs. Seven of the eight comments are six words or shorter. Four involve the unknown category on one side, either a fragment the ensemble labeled that the annotators could not or the reverse, and one of these is a Hindi--English comment of the kind discussed in the Limitations. The ensemble assigns schadenfreude where the annotators did not three times (once each against hostility, sympathy, and unknown) and misses it once, so it labels 49 of the 200 pairs as schadenfreude against 47 in the gold set.

The same pairs show how a toxicity label maps onto the human categories. Of the 47 pairs the annotators labeled schadenfreude, the toxicity classifier labels 24 toxic and 18 neutral, missing close to half of human-identified schadenfreude; of the 67 pairs it labels toxic, 40 are hostility without pleasure and 24 schadenfreude by the majority label. At the human level, toxicity therefore tracks hostility more closely than schadenfreude, the conflation that Section~4.6 documents at corpus scale.

We correct label shares for classifier error by inverting the ensemble's confusion matrix on the gold set (Section~3.3), with 95\% intervals from 2{,}000 bootstrap resamples of the gold pairs (Table~\ref{tab:adjusted}). Correction lowers the schadenfreude share by about one point in every country (overall from 26.1\% to 25.1\%) and raises hostility by 1.3--1.8 points. The three countries' adjusted schadenfreude shares lie within 0.5 points of each other, and India's lead in hostility over the other two countries widens slightly.

Post-level labels were validated on the 180-post gold set. Two annotators labeled every post, and the third labeled 59 of the first 60 posts in the set's random order, which cover all nine outlets and all ten years. On these 59 posts the gold label is the majority of three (no post had a three-way split on any field), and Table~\ref{tab:gold} reports Qwen's accuracy against it. Annotator agreement on these posts is highest for topic (Fleiss' $\kappa = 0.90$), 0.75 for target and event class, and lowest for deservedness (0.65), the field that asks for a judgment of the post's framing. Over all 180 posts, using the majority where three labels exist and the shared label where the two full-set annotators agree, accuracy is 97.6--100\% across fields. We treat these figures as secondary, because they exclude the 7 to 31 posts per field on which those two annotators disagree, which are the hardest cases.

\begin{table}[t]
\centering\small
\caption{Accuracy against the human majority label. Top: comment labels on the 200-pair gold set. Bottom: Qwen3-32B post labels on the 59 triple-annotated posts. Wilson 95\% confidence intervals in brackets.}
\label{tab:gold}
\begin{tabular}{lrr}
\toprule
 & Accuracy (\%) & Macro-F1 \\
\midrule
\multicolumn{3}{l}{\emph{Comment label ($n = 200$)}} \\
Ensemble (majority vote) & 96.0 [92.3, 98.0] & 0.929 \\
Gemma-3-27B & 87.5 [82.2, 91.4] & 0.784 \\
Qwen3-32B & 84.5 [78.8, 88.9] & 0.732 \\
Llama-3.3-70B & 84.0 [78.3, 88.4] & 0.695 \\
\midrule
\multicolumn{3}{l}{\emph{Post field, Qwen3-32B ($n = 59$)}} \\
Topic & 100.0 [93.9, 100.0] & 1.000 \\
Target & 98.3 [91.0, 99.7] & 0.973 \\
Deservedness & 94.9 [86.1, 98.3] & 0.960 \\
Event class & 94.9 [86.1, 98.3] & 0.942 \\
\bottomrule
\end{tabular}
\end{table}

\begin{table*}[t]
\centering\small
\caption{Error-adjusted comment-level label distribution (\%), overall and by country, from confusion-matrix inversion on the 200-pair gold set; 95\% bootstrap intervals (2{,}000 resamples of the gold pairs) in brackets. Raw shares are in Table~\ref{tab:prevalence}.}
\label{tab:adjusted}
\begin{tabular}{lrrrrr}
\toprule
 & SCH & HOS & SYM & NEU & UNK \\
\midrule
All comments & 25.1 [22.3, 27.2] & 32.7 [31.2, 34.8] & 15.5 [14.4, 17.8] & 22.3 [19.9, 23.8] & 4.3 [1.7, 7.9] \\
United States & 25.2 [22.5, 27.3] & 33.2 [31.8, 35.4] & 15.8 [14.7, 18.2] & 21.9 [19.6, 23.3] & 3.8 [1.2, 7.2] \\
United Kingdom & 25.0 [22.1, 27.1] & 28.8 [27.5, 30.6] & 15.5 [14.3, 17.7] & 25.5 [22.8, 27.0] & 5.3 [2.7, 9.4] \\
India & 24.7 [21.8, 26.8] & 39.3 [37.5, 41.9] & 13.9 [12.9, 16.0] & 17.2 [14.6, 18.9] & 4.8 [2.2, 8.7] \\
\bottomrule
\end{tabular}
\end{table*}

\begin{table*}[t]
\centering\small
\caption{United States power regressions ($n = 7{,}791$ post cells; $6{,}610$ dropping transition years). Outcome is the per-post share of comments labeled SCH (schadenfreude) or HOS (hostility without pleasure); cells are posts from NBC News (left) and Fox News (right) with topic labels. In-power indicates the audience's aligned party holds the presidency, coded to inauguration day where post dates exist. All models include topic fixed effects; HC1 standard errors; 95\% confidence intervals in brackets. Weighted models weight by comments per post. The transition-year specification drops 2017 and 2021; the majority-of-year specification assigns each post to the party holding office for most of its calendar year.}
\label{tab:regression}
\begin{tabular}{llrrr}
\toprule
Outcome & Specification & Right-leaning & In power & In power $\times$ Right \\
\midrule
SCH & Primary, unweighted & $+0.028$ $[+0.016, +0.040]$ & $+0.011$ $[-0.001, +0.024]$ & $-0.018$ $[-0.035, -0.002]$ \\
SCH & Primary, weighted & $+0.030$ $[+0.007, +0.052]$ & $+0.010$ $[-0.015, +0.035]$ & $-0.032$ $[-0.060, -0.003]$ \\
SCH & Majority-of-year coding & $+0.028$ $[+0.016, +0.040]$ & $+0.012$ $[-0.000, +0.024]$ & $-0.019$ $[-0.036, -0.003]$ \\
SCH & Drop transition years & $+0.009$ $[-0.006, +0.024]$ & $-0.006$ $[-0.021, +0.010]$ & $-0.003$ $[-0.023, +0.016]$ \\
HOS & Primary, unweighted & $-0.050$ $[-0.064, -0.036]$ & $-0.055$ $[-0.069, -0.041]$ & $+0.063$ $[+0.045, +0.081]$ \\
HOS & Primary, weighted & $-0.063$ $[-0.091, -0.036]$ & $-0.063$ $[-0.092, -0.034]$ & $+0.058$ $[+0.027, +0.089]$ \\
HOS & Drop transition years & $-0.029$ $[-0.046, -0.012]$ & $-0.038$ $[-0.055, -0.021]$ & $+0.040$ $[+0.019, +0.061]$ \\
\bottomrule
\end{tabular}
\end{table*}

\section{Post-Level Labeling}\label{app:topics}
Each unique post was labeled with single deterministic passes of Qwen3-32B for topic (the eight domains of Section 3.4 plus Other), the target of the misfortune (ordinary person, public figure, organization, group), the deservedness framing of the post (deserved, undeserved, none), and the event class (death, scandal, defeat, accident, hardship). Parse failures were below 0.2\% for every field; posts labeled Other for topic (4.6\%) are excluded from topic analyses. Validation uses a post-level gold set of 180 posts (20 per outlet, spread across all ten years) labeled for all four fields by two annotators in full and by the third on 59 posts; accuracy against the majority label on those 59 posts is reported in Appendix~\ref{app:validation}.

\section{A Fourth UK Outlet}\label{app:metro}
\begin{figure}[h]
\centering
\includegraphics[width=0.85\linewidth]{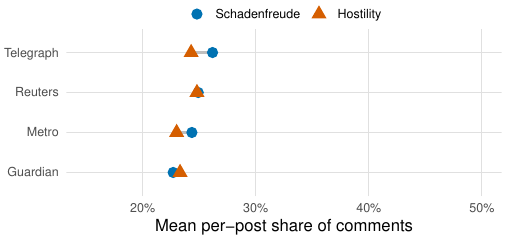}
\caption{Adding a fourth UK outlet leaves the UK results essentially unchanged. Mean per-post schadenfreude (circles) and hostility (triangles) shares for the three main UK outlets and Metro, on the same scale as Figure~\ref{fig:outlets}.}
\label{fig:metro}
\end{figure}

The main analysis uses one outlet per country--leaning cell. To check that the UK results do not depend on which left-of-center outlet occupies that cell, we collected a tenth outlet, Metro (3{,}687 misfortune posts, 99{,}244 comments), rated Left-Center by Media Bias/Fact Check, and classified it with the same pipeline. Metro's audience produces schadenfreude in 24.4\% of comments per post and hostility in 23.0\%, both inside the range of the three main UK outlets (Figure~\ref{fig:metro}). Adding Metro to the UK aggregate moves the country's mean schadenfreude rate from 24.5\% to 24.5\% and its hostility rate from 24.1\% to 23.6\%, changes smaller than the gap between any two UK outlets. Metro enters no main-text figure, table, or model.

\section{Data Transparency}\label{app:transparency}
Table~\ref{tab:transparency} reports comments per outlet and year and total misfortune posts per outlet, including the Metro robustness set of Appendix~\ref{app:metro}. Coverage is uneven in ways that reflect the outlets' posting and moderation histories rather than our sampling: OpIndia and Metro have no 2017 comments, Reuters is thin in 2017 and 2020, and Fox News comments carry no user identifiers.
Median comments per post range from 4 (OpIndia) to 48 (NBC).

\begin{table*}[t]
\centering\small
\caption{Comments per outlet and year, and total distinct misfortune posts per outlet (the nine main outlets sum to 20{,}661). A post whose comments span two calendar years contributes to both year columns but is counted once in Posts. $^\dagger$Metro is used only for the robustness check of Appendix~\ref{app:metro} and enters no main-text result.}
\label{tab:transparency}
\begin{tabular}{lllrrrrrrrrrrr}
\toprule
Country & Leaning & Outlet & Posts & 2015 & 2016 & 2017 & 2018 & 2019 & 2020 & 2021 & 2022 & 2023 & 2024 \\
\midrule
India & Left & Hindustan Times & 1460 & 8499 & 3885 & 7627 & 6161 & 2812 & 4518 & 3082 & 2107 & 1400 & 251 \\
India & Center & Indian Express & 2003 & 4646 & 7613 & 3966 & 4241 & 7278 & 5107 & 4053 & 5598 & 580 & 324 \\
India & Right & OpIndia & 2793 & 60 & 749 & 0 & 1311 & 4020 & 5036 & 5423 & 6409 & 2002 & 1239 \\
UK & Left & Guardian & 2200 & 26083 & 18142 & 22218 & 6510 & 15084 & 25180 & 20880 & 7730 & 9445 & 3480 \\
UK & Center & Reuters & 1505 & 9939 & 5200 & 61 & 432 & 1633 & 66 & 13400 & 7178 & 1574 & 1642 \\
UK & Right & Telegraph & 2049 & 13150 & 8840 & 10518 & 5910 & 9696 & 11138 & 5355 & 8013 & 2458 & 4769 \\
UK & Left-c.$^\dagger$ & Metro & 3687 & 2940 & 3315 & 0 & 15978 & 4361 & 5025 & 43622 & 3314 & 18498 & 2191 \\
USA & Left & NBC News & 3317 & 111565 & 88858 & 29043 & 35328 & 12025 & 23429 & 63785 & 66408 & 3004 & 16728 \\
USA & Center & CNBC & 719 & 1371 & 673 & 2003 & 1895 & 3089 & 8212 & 9528 & 437 & 373 & 1385 \\
USA & Right & Fox News & 4615 & 37518 & 40180 & 6214 & 13880 & 13532 & 19989 & 2838 & 24215 & 24414 & 17122 \\
\bottomrule
\end{tabular}
\end{table*}

\section{Ethical Considerations}\label{app:ethics}
The study analyzes public content from the public pages of news organizations; no private or restricted content was accessed and no users were contacted. We report only aggregates, remove identifying information before analysis, and never analyze, profile, or report individual users; the example comments in Appendix~\ref{app:examples} are paraphrased so that no verbatim comment is searchable, and heavy commenters are examined only in aggregate. Because the study observes public online behavior rather than interacting with human subjects, it falls outside institutional review requirements for human-subjects research; we nonetheless follow established practice for social media data, including data minimization, aggregate-only reporting, and suppression of small cells. Annotators consented to the use of their labels in this research.

\end{document}